# Low-Complexity Robust Beamforming Design for IRS-Aided MISO Systems with Imperfect Channels

Yasaman Omid, Seyyed MohammadMahdi Shahabi, Cunhua Pan, Yansha Deng, Arumugam Nallanathan, *Fellow, IEEE*

*Abstract*—In this paper, large-scale intelligent reflecting surface (IRS)-assisted multiple-input single-output (MISO) system is considered in the presence of channel uncertainty. To maximize the average sum rate of the system by jointly optimizing the active beamforming at the BS and the passive phase shifts at the IRS, while satisfying the power constraints, a novel robust beamforming design is proposed by using the penalty dual decomposition (PDD) algorithm. By applying the upper bound maximization/minimization (BSUM) method, in each iteration of the algorithm, the optimal solution for each variable can be obtained with closed-form expression. Simulation results show that the proposed scheme achieves high performance with very low computational complexity.

## I. INTRODUCTION

INTELLIGENT reflecting surface (IRS) is one of the most promising techniques to improve the spectral efficiency in wireless communication systems. The basic function of the IRS is to reconfigure the wireless propagation environment using a meta-surface with a number of artificial passive elements [1], [2]. To be specific, by designing a passive beamforming at the IRS, the electromagnetic waves could be reflected towards a specific direction. Due to the passive nature of the IRS, channel acquisition in IRS-aided systems seems to be challenging, especially for the IRS-related channels. The reason is that the IRS does not possess active radio frequency chain or any other signal processing units for sending pilot signals or processing training signals, respectively. To address this issue, in [3], the authors installed several active elements in IRS for channel estimation. However, this causes extra power consumption and data exchange overhead for the system. Fortunately, it is found that the cascaded BS-IRS-user channel is sufficient for the transmission design. Hence, most of the existing works focused on the study of the estimation of BS-IRS-user channels [4]. Although the channel estimation for IRS-aided communication systems has been extensively studied, most of the existing transmission design is based on the idealistic assumption of perfect channel state information (CSI). There are only a paucity of research works that have studied the robust transmission design by considering the imperfect CSI [5]–[7].

The robust transmission design for IRS-aided systems was first studied in [5], where the authors considered a bounded CSI error model for an IRS-aided multi-user MISO system. By resorting to the technique of semidefinite programming (SDP), the transmit power minimization problem was converted into a sequence of convex sub-problems that can be efficiently solved. In [6], the transmit power minimization problem for an IRS-aided MISO system was formulated subject to the worst-case rate constraint under the bounded cascaded channel error model, and subject to the outage probability constraints under statistical cascaded channel error model. Low-complexity but efficient algorithms were proposed to handle these problems. In [7] a new robust algorithm for joint BS-IRS beamforming design is presented based on mean square error (MSE) minimization for an IRS-aided single-user MISO system. To tackle this problem, they adopted alternating algorithm (AO) based on the majorization-minimization (MM) technique and the Lagrangian dual method, in which closed-form solution for each variable can be derived in each iteration. In both [5] and [6], the proposed algorithms were based on the iterative procedure, where each iteration needs to solve convex optimization problems via the CVX tool, which incurs excessive computational complexity. Although the iterative algorithm in [7] can obtain closed-form solution in each iteration, the method is only applicable for single-user systems.

Motivated by above, in this paper, a new algorithm for joint BS and IRS beamforming design is devised by utilzing the penalty dual decomposition (PDD) technique [8], [9], which readily leads to a less complex structure compared to the MM-driven method. We assume that the BS only possesses imperfect CSI, and an statistical CSI error model is considered for the cascaded BS-IRS-user and BS-user channels. First, we formulate a rate maximization problem subject to power constraints. To facilitate the algorithm design, by using the block successive upper bound maximization/minimization (BSUM) method [10], we reformulate this optimization problem into a tractable manner. Through the PDD algorithm, the problem is solved iteratively, and the closed-form solution for each variable is obtained in each iteration. Based on the complexity analysis, it can be concluded that the complexity of the proposed solution is lower than that of the existing methods.

Yasaman Omid, Cunhua Pan and Arumugam Nallanathan are with the School of Electronic Engineering and Computer Science, Queen Mary University of London, U.K. (e-mail: y.omid@qmul.ac.uk; c.pan@qmul.ac.uk; a.nallanathan@qmul.ac.uk).

Seyyed MohammadMahdi Shahabi is with the Department of Electrical Engineering, K. N. Toosi University of Technology, Tehran, Iran (e-mail: shahabi@ee.kntu.ac.ir).

Yansha Deng is with the Department of Engineering, Kings College London, U.K. (e-mail: yansha.deng@kcl.ac.uk).

## II. System Model and Problem Formulation

We consider the downlink IRS-aided transmission of a multi-antenna base station (BS) with $N_T$ antennas serving $K$ single-antenna users. The IRS has $M$ phase shifters (PS). The BS-IRS, BS-user and IRS-user channel matrices are denoted by $\mathbf{G}_{BI} \in \mathbb{C}^{M \times N_T}$, $\mathbf{G}_{BU} \in \mathbb{C}^{K \times N_T}$, and $\mathbf{G}_{IU} \in \mathbb{C}^{K \times M}$, respectively. We consider an indoor application, where there is rich scattering; hence, we consider only non line of sight channels in our model. This assumption has been considered for IRS-related links in many papers such as [11]. Each row of the channel matrices $\mathbf{G}_{BU}$ and $\mathbf{G}_{IU}$ denotes the corresponding channel vectors of the $i$-th user, and follows zero mean complex Gaussian distribution as $\mathbf{g}_{BU}^{[i]} \sim \mathcal{CN}(0, \beta_{BU}^{[i]}\mathbf{I})$ and $\mathbf{g}_{IU}^{[i]} \sim \mathcal{CN}(0, \beta_{IU}^{[i]}\mathbf{I})$, whereas $\text{vec}(\mathbf{G}_{BI}) \sim \mathcal{CN}(0, \beta_{BI}\mathbf{I})$. Note that $\beta_{BU}^{[i]}$, $\beta_{IU}^{[i]}$ and $\beta_{BI}$ represent the large scale fading gains of the corresponding channels, modeled by the 3GPP standards. Considering non-perfect channel estimation for the BS-user and the IRS-user channels, the estimated channel matrices are represented by $\hat{\mathbf{G}}_{BU}$ and $\hat{\mathbf{G}}_{IU}$ respectively while $\tilde{\mathbf{G}}_{BU}$ and $\tilde{\mathbf{G}}_{IU}$ stand for the estimation error matrices of these channels. To be more specific, we have $\mathbf{G}_{BU} = \hat{\mathbf{G}}_{BU} + \tilde{\mathbf{G}}_{BU}$ and $\mathbf{G}_{IU} = \hat{\mathbf{G}}_{IU} + \tilde{\mathbf{G}}_{IU}$. Note that the channel estimation errors is assumed to have complex normal distribution as $\text{vec}(\tilde{\mathbf{G}}_{BU}) \sim \mathcal{CN}(0, \sigma_{BU}^2\mathbf{I})$ and $\text{vec}(\tilde{\mathbf{G}}_{IU}) \sim \mathcal{CN}(0, \sigma_{IU}^2\mathbf{I})$. The received signal at the $i$-th user can be formulated as

$$y^{[i]} = \sum_{k=1}^{K}\left[\mathbf{g}_{BU}^{[i]}\mathbf{v}_k s^{[k]} + \mathbf{g}_{IU}^{[i]}\mathbf{\Psi}_I\mathbf{G}_{BI}\mathbf{v}_k s^{[k]}\right] + w^{[i]}$$
$$= \left(\mathbf{g}_{BU}^{[i]} + \mathbf{g}_{IU}^{[i]}\mathbf{\Psi}_I\mathbf{G}_{BI}\right)\mathbf{v}_i s^{[i]}$$
$$\quad + \sum_{k \neq i}^{K}\left[\left(\mathbf{g}_{BU}^{[i]} + \mathbf{g}_{IU}^{[i]}\mathbf{\Psi}_I\mathbf{G}_{BI}\right)\mathbf{v}_k s^{[k]}\right] + w^{[i]}$$
$$= \mathbf{g}^{[i]}\mathbf{v}_i s^{[i]} + \sum_{k \neq i}^{K}\left[\mathbf{g}^{[i]}\mathbf{v}_k s^{[k]}\right] + w^{[i]} \quad (1)$$

where $\mathbf{g}^{[i]} = \mathbf{g}_{BU}^{[i]} + \mathbf{g}_{IU}^{[i]}\mathbf{\Psi}_I\mathbf{G}_{BI}$ denotes the effective channel between the BS and the $i$-th user, $\mathbf{V} = [\mathbf{v}_1, \ldots, \mathbf{v}_K]$ is the BS beamforming matrix, and $\mathbf{s}$ is the desired symbol vector, whose elements have unit power, i.e., $\mathbb{E}\left\{|s^{[i]}|^2\right\} = 1$. Also, $\mathbf{\Psi}_I = \text{diag}[\psi_1, \ldots, \psi_M]$ is the diagonal phase shift matrix in the IRS and $w^{[i]} \sim \mathcal{CN}(0, \sigma_w^2)$ is the additive white Gaussian noise at the $i$th user. The received signal in (1) can be rewritten in terms of the estimated channel vectors as

$$y^{[i]} = \left(\hat{\mathbf{g}}_{BU}^{[i]} + \hat{\mathbf{g}}_{IU}^{[i]}\mathbf{\Psi}_I\mathbf{G}_{BI}\right)\mathbf{v}_i s^{[i]}$$
$$+ \sum_{k \neq i}^{K}\left[\left(\hat{\mathbf{g}}_{BU}^{[i]} + \hat{\mathbf{g}}_{IU}^{[i]}\mathbf{\Psi}_I\mathbf{G}_{BI}\right)\mathbf{v}_k s^{[k]}\right]$$
$$+ \sum_{k=1}^{K}\left[\left(\tilde{\mathbf{g}}_{BU}^{[i]} + \tilde{\mathbf{g}}_{IU}^{[i]}\mathbf{\Psi}_I\mathbf{G}_{BI}\right)\mathbf{v}_k s^{[k]}\right] + w^{[i]}$$
$$= \hat{\mathbf{g}}^{[i]}\mathbf{t}^{[i]} + \sum_{k \neq i}^{K}\left[\hat{\mathbf{g}}^{[i]}\mathbf{t}^{[k]}\right] + \sum_{k=1}^{K}\left[\tilde{\mathbf{g}}^{[i]}\mathbf{t}^{[k]}\right] + w^{[i]}, \quad (2)$$

where $\mathbf{t}^{[i]} \triangleq \mathbf{v}_i s^{[i]}$ denotes the transmit vector intended for the $i$-th user, the terms $\hat{\mathbf{g}}_{BU}^{[i]}$, $\hat{\mathbf{g}}_{IU}^{[i]}$, $\tilde{\mathbf{g}}_{BU}^{[i]}$ and $\tilde{\mathbf{g}}_{IU}^{[i]}$ stand for the $i$th row of the matrices $\hat{\mathbf{G}}_{BU}$, $\hat{\mathbf{G}}_{IU}$, $\tilde{\mathbf{G}}_{BU}$ and $\tilde{\mathbf{G}}_{IU}$ respectively, and we have

$$\hat{\mathbf{g}}^{[i]} \triangleq \hat{\mathbf{g}}_{BU}^{[i]} + \mathbf{f}\hat{\mathbf{G}}_c^{[i]}, \quad (3)$$

$$\tilde{\mathbf{g}}^{[i]} \triangleq \tilde{\mathbf{g}}_{BU}^{[i]} + \mathbf{f}\tilde{\mathbf{G}}_c^{[i]}, \quad (4)$$

in which $\mathbf{f} \triangleq (\text{diag}(\mathbf{\Psi}_I))^T = [\psi_1, \ldots, \psi_M]$, $\hat{\mathbf{G}}_c^{[i]} = \text{diag}(\hat{\mathbf{g}}_{IU}^{[i]})\mathbf{G}_{BI}$ is the estimated cascaded channel and $\tilde{\mathbf{G}}_c^{[i]} = \text{diag}(\tilde{\mathbf{g}}_{IU}^{[i]})\mathbf{G}_{BI}$ is the cascaded channel estimation error. For large-scale IRS with a large number of PSs (e.g., $M \to \infty$), the distribution of $\tilde{\mathbf{g}}^{[i]}$ is

$$\tilde{\mathbf{g}}^{[i]} \sim \mathcal{CN}\left(0, \left(\sigma_{BU}^2 + \sigma_{IU}^2 \beta_{BI} M\right)\mathbf{I}\right) = \mathcal{CN}\left(0, \sigma_g^2 \mathbf{I}\right). \quad (5)$$

The detailed steps of deriving the distribution of $\tilde{\mathbf{g}}^{[i]}$ can be found in Appendix A. The approximation in (5) becomes more accurate when the value of $M$ increases. Note that, the accuracy of this approximation does not depend on the value of $N_T$, hence the design is suitable for any number of transmit antennas at the BS. Additionally, the effective estimated channel matrix is represented by

$$\hat{\mathbf{G}} \triangleq \left[\hat{\mathbf{g}}^{[1]\,T}, \ldots, \hat{\mathbf{g}}^{[K]\,T}\right]^T = \mathbf{G} + \tilde{\mathbf{G}}, \quad (6)$$

in which we have $\mathbf{G} \triangleq \left[\mathbf{g}^{[1]\,T}, \ldots, \mathbf{g}^{[K]\,T}\right]^T$ and $\tilde{\mathbf{G}} \triangleq \left[\tilde{\mathbf{g}}^{[1]\,T}, \ldots, \tilde{\mathbf{g}}^{[K]\,T}\right]^T$. Considering all the above, the following theorem is presented.

**Theorem 1.** *It can be shown that the minimum achievable rate of the $i$-th user is given by*

$$R^{[i]} = \log_2\left(1 + \frac{|\hat{\mathbf{g}}^{[i]}\mathbf{v}_i|^2}{\sum_{\substack{k=1\\k\neq i}}^{K}[|\hat{\mathbf{g}}^{[i]}\mathbf{v}_k|^2] + \sigma_g^2 \sum_{k=1}^{K}[|\mathbf{v}_k|^2] + \sigma_w^2}\right). \quad (7)$$

*Proof.* Let $\mathcal{I}\left(\mathbf{t}^{[i]}; y^{[i]}|\hat{\mathbf{G}}\right)$ be the conditional mutual information of user $i$ conditioned on estimated channel matrix $\hat{\mathbf{G}}$. Expanding $\mathcal{I}\left(\mathbf{t}^{[i]}; y^{[i]}|\hat{\mathbf{G}}\right)$ in terms of the differential entropies results in

$$\mathcal{I}\left(\mathbf{t}^{[i]}; y^{[i]}|\hat{\mathbf{G}}\right) = \mathcal{H}\left(\mathbf{t}^{[i]}|\hat{\mathbf{G}}\right) - \mathcal{H}\left(\mathbf{t}^{[i]}|y^{[i]}, \hat{\mathbf{G}}\right). \quad (8)$$

The first term on the right hand side of (8) simplifies to $\log_2 \det(2\pi e \mathbf{F}_i)$, where $\mathbf{F}_i \triangleq \mathbb{E}\left\{\mathbf{t}^{[i]}\mathbf{t}^{[i]H}\right\}$ denotes the transmit covariance matrix related to $\mathbf{t}^{[i]}$ [12]. Regarding the equation in (2), the second term of the right hand side of (8) is upper bounded by the entropy of a Gaussian random variable [13] as follows

$$\mathcal{H}\left(\mathbf{t}^{[i]}|y^{[i]}, \hat{\mathbf{G}}\right) \leq \log_2 \det\left(2\pi e \left(\mathbf{F}_i - \frac{\mathbf{F}_i \hat{\mathbf{g}}^{[i]H}\hat{\mathbf{g}}^{[i]}\mathbf{F}_i}{\hat{\mathbf{g}}^{[i]}\mathbf{F}_i \hat{\mathbf{g}}^{[i]H} + \Gamma_i}\right)\right), \quad (9)$$

in which $\Gamma_i \triangleq \sum_{\substack{k=1\\k\neq i}}^{K} \hat{\mathbf{g}}^{[i]H}\mathbf{F}_k \hat{\mathbf{g}}^{[i]} + \sum_{k=1}^{K} \sigma_g^2 \text{Tr}(\mathbf{F}_k) + \sigma_w^2$. Now, we employ the Woodbury matrix identity as follows

$$(\mathbf{A} + \mathbf{B}\mathbf{C}\mathbf{D})^{-1} = \mathbf{A}^{-1} - \mathbf{A}^{-1}\mathbf{B}\left(\mathbf{C}^{-1} + \mathbf{D}\mathbf{A}^{-1}\mathbf{B}\right)^{-1}\mathbf{D}\mathbf{A}^{-1}. \quad (10)$$

Assuming $\mathbf{A} = \mathbf{I}$, $\mathbf{B} = \hat{\mathbf{g}}^{[i]H}$, $\mathbf{C} = \Gamma_i^{-1}$ and $\mathbf{D} = \mathbf{g}^{[i]}\mathbf{F}_i$, and using (10), we have

$$\mathbf{I} - \frac{\hat{\mathbf{g}}^{[i]H}\hat{\mathbf{g}}^{[i]}\mathbf{F}_i}{\hat{\mathbf{g}}^{[i]}\mathbf{F}_i\hat{\mathbf{g}}^{[i]H} + \Gamma_i} = \left(\mathbf{I} + \frac{\hat{\mathbf{g}}^{[i]H}\hat{\mathbf{g}}^{[i]}\mathbf{F}_i}{\Gamma_i}\right)^{-1}. \quad (11)$$

Thus, the right hand side of (9) can be rewritten as

$$\log_2 \det \left(2\pi e \mathbf{F}_i \left(\mathbf{I} - \frac{\hat{\mathbf{g}}^{[i]H}\hat{\mathbf{g}}^{[i]}\mathbf{F}_i}{\hat{\mathbf{g}}^{[i]}\mathbf{F}_i\hat{\mathbf{g}}^{[i]H} + \Gamma_i}\right)\right)$$
$$= \log_2 \det \left(2\pi e \mathbf{F}_i \left(\mathbf{I} + \frac{\hat{\mathbf{g}}^{[i]H}\hat{\mathbf{g}}^{[i]}\mathbf{F}_i}{\Gamma_i}\right)^{-1}\right)$$
$$= \log_2 \det (2\pi e \mathbf{F}_i) - \log_2 \det \left(\mathbf{I} + \frac{\hat{\mathbf{g}}^{[i]H}\hat{\mathbf{g}}^{[i]}\mathbf{F}_i}{\Gamma_i}\right). \quad (12)$$

Exploiting (12) and employing Sylvester's determinant theorem, i.e., $\det(\mathbf{I} + \mathbf{AB}) = \det(\mathbf{I} + \mathbf{BA})$, (9) is rewritten as

$$\mathcal{H}\left(\mathbf{t}^{[i]}|y^{[i]}, \hat{\mathbf{G}}\right) \leq \log_2 \det \left(2\pi e \mathbf{F}_i \left(\mathbf{I} - \frac{\hat{\mathbf{g}}^{[i]H}\hat{\mathbf{g}}^{[i]}\mathbf{F}_i}{\hat{\mathbf{g}}^{[i]}\mathbf{F}_i\hat{\mathbf{g}}^{[i]H} + \Gamma_i}\right)\right)$$
$$= \log_2 \det (2\pi e \mathbf{F}_i) - \log_2 \det \left(1 + \frac{\hat{\mathbf{g}}^{[i]}\mathbf{F}_i\hat{\mathbf{g}}^{[i]H}}{\Gamma_i}\right). \quad (13)$$

Consequently, assuming $\mathbb{E}\left\{|s^{[i]}|^2\right\} = 1$ and utilizing (13), (8), we derive

$$\mathcal{I}\left(\mathbf{t}^{[i]}; y^{[i]}|\hat{\mathbf{G}}\right) \geq \log_2 \det \left(1 + \frac{\hat{\mathbf{g}}^{[i]}\mathbf{F}_i\hat{\mathbf{g}}^{[i]H}}{\Gamma_i}\right)$$
$$= \log_2 \left(1 + \frac{|\hat{\mathbf{g}}^{[i]}\mathbf{t}_i|^2}{\Gamma_i}\right) = \log_2 \left(1 + \frac{|\hat{\mathbf{g}}^{[i]}\mathbf{v}_i|^2}{\Gamma_i}\right). \quad (14)$$

Therefore, the minimum achievable rate of the $i$-th user is written as (7) and thus the proof is completed. □

Now, in order to jointly design efficient beamforming matrices for the BS and the IRS, the sum achievable rate in (7) is maximized. To this end, we formulate the following weighted sum rate maximization problem

$$\max_{\mathbf{V},\mathbf{f}} \quad \sum_{i=1}^{K} \alpha_i R^{[i]}$$
$$\text{s.t.} \quad |\psi_i| = 1, i = 1, ..., M, \quad (15)$$
$$\text{Tr}\left(\mathbf{V}\mathbf{V}^H\right) \leq P_T,$$

where $\alpha_i$ denotes the weight measuring the priority of the $i$th user and $P_T$ stands for the BS power budget. It is readily seen that, the optimization problem in (15) is challenging to solve, as it is a non-convex problem with multiple coupled optimization variables. In the following, we aim to find a tractable low-complexity solution for this problem.

## III. PDD-BASED SOLUTION

In this section, we aim to provide a tractable low-complexity solution to the optimization problem in (15). To this end, the PDD method is utilized. We first introduce a set of auxiliary variables $\{\mathbf{X}, \bar{\mathbf{V}}\}$ and we define $\mathcal{X} = \{\mathbf{f}, \mathbf{V}, \bar{\mathbf{V}}, \mathbf{X}\}$. In order to use the PDD method for our problem, we need to be able to decouple the optimization variables. To do so, we rewrite the optimization problem in (15) as

$$\max_{\mathcal{X}} \quad \sum_{i=1}^{K} \alpha_i \log_2 \left(1 + \frac{|x_{ii}|^2}{\sum_{\substack{k=1 \\ k \neq i}}^{K} |x_{ki}|^2 + \sigma_g^2 \sum_{k=1}^{K} |\mathbf{v}_k|^2 + \sigma_w^2}\right)$$
$$\text{s.t.} \quad \text{Tr}\left(\bar{\mathbf{V}}\bar{\mathbf{V}}^H\right) \leq P_T$$
$$|\psi_i| = 1, i = 1, ..., M,$$
$$\bar{\mathbf{V}} = \mathbf{V}, \mathbf{X} = \mathbf{V}^H\hat{\mathbf{G}}^H,$$
$$(16)$$

where the term $x_{ij}$ denotes the element in the $i$-th row and the $j$-th column of the matrix $\mathbf{X}$. By appending all of the equality constraints to the objective function, the Augmented Lagrangian (AL) problem can be obtained as follows

$$\max_{\mathcal{X}} \quad \sum_{i=1}^{K} \alpha_i \log_2 \left(1 + \frac{|x_{ii}|^2}{\sum_{\substack{k=1 \\ k \neq i}}^{K} |x_{ki}|^2 + \sigma_g^2 \sum_{k=1}^{K} |\mathbf{v}_k|^2 + \sigma_w^2}\right)$$
$$- P_\rho(\mathcal{X})$$
$$\text{s.t.} \quad \text{Tr}\left(\bar{\mathbf{V}}\bar{\mathbf{V}}^H\right) \leq P_T,$$
$$|\psi_i| = 1, i = 1, ..., M,$$
$$(17)$$

in which the function $P_\rho(\mathcal{X})$ is defined as $P_\rho(\mathcal{X}) = \frac{1}{2\rho}\left(||\mathbf{V} - \bar{\mathbf{V}} + \rho\mathbf{Z}_v||^2 + ||\mathbf{X} - \mathbf{V}^H\hat{\mathbf{G}}^H + \rho\mathbf{Z}_g||^2\right)$. The variable $\rho$ denotes the penalty parameter of the Lagrangian function and the matrices $\mathbf{Z}_v$ and $\mathbf{Z}_g$ are the dual variables associated with their respective equality constraint. Now, the key to use the BSUM method is to find a tractable locally lower bound for the objective function of the problem in (15). To do so, by employing the theory of the WMMSE method, the following theorem is achieved.

**Theorem 2.** *For each user $i$ the following inequality holds*

$$\log_2 \left(1 + \frac{|x_{ii}|^2}{\sum_{k \neq i}^{K} |x_{ki}|^2 + \sigma_g^2 \sum_{k=1}^{K} |\mathbf{v}_k|^2 + \sigma_w^2}\right) \geq$$
$$\log_2(w_i) - w_i e_i(u_i, \mathbf{X}, \mathbf{V}) + 1, \quad (18)$$

where $e_i(u_i, \mathbf{X}, \mathbf{V}) = |1 - u_i^* x_{ii}|^2 + \sum_{k \neq i} |u_i^* x_{ki}|^2 + \sigma_g^2 \sum_{k=1}^{K} ||u_i^* \mathbf{v}_k||^2 + \sigma_w^2 |u_i|^2$.

The proof can be straightforwardly provided by using the first-order optimality condition for the right hand side of the inequality. In this case, the optimal values for $u_i$ and $w_i$ are calculated by

$$u_i^{opt} = \frac{x_{ii}}{\sum_{k=1}^{K} |x_{ki}|^2 + \sigma_g^2 \sum_{k=1}^{K} ||\mathbf{v}_k||^2 + \sigma_w^2}, \quad (19)$$

$$w_i^{opt} = 1 + \frac{|x_{ii}|^2}{\sum_{k \neq i}^{K} |x_{ki}|^2 + \sigma_g^2 \sum_{k=1}^{K} ||\mathbf{v}_k||^2 + \sigma_w^2}. \quad (20)$$

It can be readily proved that the right hand side of the inequality in (18) is a tight lower bound for the objective function of (16). Now by replacing the objective function with its tractable lower bound, (17) can be rewritten as

$$\min_{\mathcal{X}} \quad \sum_{i=1}^{K} \alpha_i w_i e_i(u_i, \mathbf{X}, \mathbf{V}) + P_\rho(\mathcal{X})$$
$$\text{s.t.} \quad \text{Tr}\left(\bar{\mathbf{V}}\bar{\mathbf{V}}^H\right) \leq P_T, \quad (21)$$
$$|\psi_i| = 1, i = 1, ..., M.$$

For the sake of simplicity, we reformulate the optimization

problem in (21) as

$$\min_{\mathcal{X}} \quad \text{Tr}(\mathbf{X}\mathbf{B}\mathbf{X}^H) - \text{Tr}(\mathbf{A}\mathbf{X}\mathbf{D}^H) - \text{Tr}(\mathbf{A}\mathbf{X}^H\mathbf{D})$$
$$+ \text{Tr}(\mathbf{A}) + \sigma_g^2 \text{Tr}(\mathbf{V}^H b\mathbf{V}) + \sigma_w^2 b + P_\rho(\mathcal{X}) \quad (22)$$
$$\text{s.t.} \quad \text{Tr}(\bar{\mathbf{V}}\bar{\mathbf{V}}^H) \leq P_T,$$
$$|\psi_i| = 1, i = 1, ..., M.$$

where $\mathbf{B} = \text{diag}\{[\alpha_1 w_1 |u_1|^2, \alpha_2 w_2 |u_2|^2, \ldots, \alpha_K w_K |u_K|^2]\}$, $b = \text{Tr}(\mathbf{B}) = \sum_{i=1}^{K} \alpha_i w_i |u_i|^2$, $\mathbf{D} = \text{diag}\{[u_1, u_2, \ldots, u_K]\}$ and $\mathbf{A} = \text{diag}\{[\alpha_1 w_1, \alpha_2 w_2, \ldots, \alpha_K w_K]\}$.

Now the remaining task is to solve the optimization problem via BSUM iterations, which consist of the following four steps. In each step, closed-form solutions are calculated for a sub-set of the optimization variables, and the BSUM steps are repeated until some convergence criteria is met.

*1)* In the first step, we solve the optimization problem (22) with respect to $\mathbf{V}$ assuming all other variables are constant. In this case, the sub-problem is an unconstrained quadratic optimization problem in which by using the first order optimality condition, we have

$$\mathbf{V} = (2\rho\sigma_g^2 b\mathbf{I} + \mathbf{I} + \hat{\mathbf{G}}^H\hat{\mathbf{G}})^{-1}(\bar{\mathbf{V}} - \rho\mathbf{Z}_v + \hat{\mathbf{G}}^H\mathbf{X}^H + \rho\hat{\mathbf{G}}^H\mathbf{Z}_g^H). \quad (23)$$

*2)* In the second step, we solve the problem for the auxiliary variable $\bar{\mathbf{V}}$. In this case, the problem becomes a projection of a point into a ball centered at the origin. This yields a closed-form solution as follows

$$\bar{\mathbf{V}} = \mathcal{P}_{P_T}(\mathbf{V} + \rho\mathbf{Z}_v), \quad (24)$$

where $\mathcal{P}_{\mathcal{Y}}(y)$ is the projection of $y$ into the convex set $\mathcal{Y}^1$.

*3)* In the third step, we solve the problem for the variable $\mathbf{X}$ assuming all other variables are constant. A closed-form solution is found by using the first order optimality condition

$$\mathbf{X} = (2\rho\mathbf{D}\mathbf{A} + \mathbf{V}^H\hat{\mathbf{G}}^H - \rho\mathbf{Z}_g)(2\rho\mathbf{B} + \mathbf{I})^{-1}. \quad (25)$$

*4)* Finally, the problem should be solved for the variable $\mathbf{f}$. In this case, the optimization problem can be reformulated as

$$\min_{|\psi_i|=1, i=1,...,M} \mathbf{f}\mathbf{H}\mathbf{f}^H - 2\text{Re}\{\mathbf{c}\mathbf{f}^H\}, \quad (26)$$

where

$$\mathbf{H} = \frac{1}{2\rho} \sum_{i=1}^{K} \sum_{j=1}^{K} \hat{\mathbf{G}}_c^{[j]} \mathbf{v}_i \mathbf{v}_i^H \hat{\mathbf{G}}_c^{[j]H}, \quad (27)$$

$$\mathbf{c} = \frac{1}{2\rho} \sum_{i=1}^{K} \sum_{j=1}^{K} \left( x_{ij}^* + \rho^* z_{g_{ij}}^* - \hat{\mathbf{g}}_{BU}^{[j]} \mathbf{v}_i \right) \left( \mathbf{v}_i^H \hat{\mathbf{G}}_c^{[j]H} \right). \quad (28)$$

With respect to only one specified $\psi_k, k = 1, ..., M$, the optimization problem becomes

$$\min_{|\psi_k|=1} |\psi_k|^2 \mathbf{H}_{kk} - 2\text{Re}\{(\mathbf{c}_k - \sum_{i \neq k}^{M} \psi_i \mathbf{H}_{ik})\psi_k^*\}, \quad (29)$$

and the solution to this optimization problem is

$$\psi_k = \frac{(\mathbf{c}_k - \sum_{i \neq k}^{M} \psi_i \mathbf{H}_{ik})}{|\mathbf{c}_k - \sum_{i \neq k}^{M} \psi_i \mathbf{H}_{ik}|}. \quad (30)$$

---
[1]The projection of a point $\mathbf{X}$ into a set $\mathbb{S}$ is defined by $\min_{\mathbf{P} \in \mathbb{S}} \|\mathbf{X} - \mathbf{P}\|$. In case $\mathbb{S}$ is a sphere centered at the origin with radius $R$ ($\mathbb{S} = \{\mathbf{X}|\|\mathbf{X}\| \leq R\}$), the projection of $\mathbf{X}$ is obtained by $R \frac{\mathbf{X}}{\|\mathbf{X}\| + \max(0, R - \|\mathbf{X}\|)}$.

Through an iterative algorithm, the optimal IRS phase shifts could be obtained. In each step, all but one of the IRS phase shifts are fixed and the problem is solved for each $\psi_k, k = 1, ..., M$. This continues until the convergence criteria is met.

The BSUM algorithm for solving the optimization problem in (22) is summarized in Algorithm 1. In addition, the PDD method, in which the dual variables and the penalty parameter are updated is provided in Table I of [8], in which an adaptive strategy is presented that switches between the penalty method and AL, and finally finds an appropriate penalty parameter $\rho$ with which the AL method could converge. A comprehensive discussion related to the convergence properties of the PDD-based algorithm is also presented in [8]. Particularly, under appropriate conditions, the sequence of $\mathbf{x}^{[i]} \in \mathcal{X}$ generated by the PDD method tends to a KKT point of the main problem.

---

**Algorithm 1** The BSUM steps to solve problem (22)

---
**Initialize** $\mathbf{f}, \mathbf{V}$ such that all constraints are met
**Compute** $\hat{\mathbf{G}}$ based on (3) and (6)
**Set** $\quad \mathbf{X} = \mathbf{V}^H \hat{\mathbf{G}}^H, \bar{\mathbf{V}} = \mathbf{V}$
**Repeat**
  1. compute $u_i^{opt}$ and $w_i^{opt}$ by (19) and (20)
  2. compute $\mathbf{V}, \bar{\mathbf{V}}$ and $\mathbf{X}$ by (23), (24) and (25)
  3. compute $\mathbf{f}$ by solving (26)
**Until** some convergence criteria are met.

---

## IV. COMPLEXITY ANALYSIS

The proposed PDD-based method is an iterative algorithm, and the computational complexity of each iteration consists of calculating $\mathbf{V}, \bar{\mathbf{V}}, \mathbf{X}$ and $\mathbf{H}$, the complexity of which are given by $\mathcal{O}(N_T^3 + KN_T^2 + K^2N_T)$, $\mathcal{O}(KN_T^2)$, $\mathcal{O}(N_TK^2 + K^3)$ and $\mathcal{O}(N_T^2M + M^2N_T)$, respectively. For a simple system with $K = 1$, $N_T = 10$ and $M = 10$, the complexity of each iteration of the presented PDD-based method is 0.1 times the complexity of the AO-based algorithm in [7], and 0.001 times the complexity of the convex algorithm presented in [5] for the full channel uncertainty case with statistical error model.

## V. NUMERICAL RESULTS

In this section, the performance of our proposed algorithm is evaluated in terms of the average sum-rate of the system. It is assumed that the large-scale fading coefficient between any two nodes is modeled by $10 \log_{10}(\beta) = -127.8 - 27 \log_{10}(d) + Z$, where $d$ denotes the distance between the two nodes in km and $Z \sim \mathcal{CN}(0, \sigma_{shad}^2)$ in which $\sigma_{shad}^2 = 8$ dB represents the shadowing. The noise variance at each user is $\sigma_w^2 = 1$, the estimation error variances are $\sigma^2 = \sigma_{BU}^2 = \sigma_{IU}^2 = 0.1$, and the priority weights of all users are $\alpha_i = 1, i = 1, \ldots K$. The initial value for the penalty parameter is set to $\rho_0 = \frac{500K}{2KM + M^2 + KN_T}$ as in section III of [9], and it is decreased by a coefficient of 0.7 in each step of the penalty method. Also, the non-robust design refers to a case with imperfect CSI where the effect of channel estimation error is not considered in the design, i.e. $\sigma_g$ in all equations of section III is set to zero. Fig. 1 illustrates the average sum rate versus the number of IRS reflectors. In this figure, the system variables are $K = 4$, $N_T = 4$, SNR= 10 dB. It is shown that the performance of the proposed robust method is close to the perfect CSI scenario. Note that, as $M$ increases, the gap between the robust design and the non-robust structure widens, which is a result of the approximation in (5), since the approximation in (5) becomes more accurate when $M$ is large. In other words, the fact that (5) is more accurate when $M$ is large, results in better performance of the robust design in larger values of $M$. Moreover, as the channel estimation error increases, the performance gain of the robust design over non-robust system becomes more obvious. In Fig. 2, the performance of the presented PDD-based method is comparable to the AO-based scheme in [7]. The system variables are $N_T = 4$ and $K = 1$. As depicted, the PDD-based algorithm outperforms the AO-based scheme in terms of average sum-rate. This should be as a result of the optimization objective in [7] which attempts to minimize the MSE.

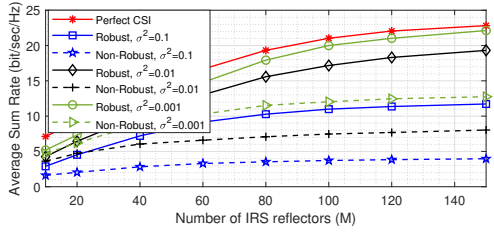

Figure 1: Average sum-rate versus $M$; A comparison among robust, non-robust and perfect CSI designs.

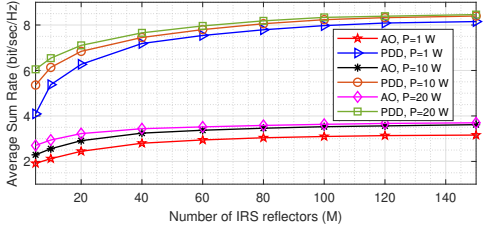

Figure 2: Average sum-rate versus $M$; A comparison between the PDD-based scheme and the AO-based scheme.

## VI. CONCLUSION

In this paper, a new robust design was presented for an IRS-assisted MISO system with channel uncertainty. The PDD algorithm is utilized to tackle the optimization problem. It is shown that the complexity of the proposed algorithm is low, and it achieves higher average sum rates over other schemes in the literature.

## APPENDIX A
## DISTRIBUTION OF $\tilde{\mathbf{g}}^{[i]}$

Here, the distribution of $\tilde{\mathbf{g}}^{[i]}$ is calculated. Based on the channel models it is obvious that $\tilde{\mathbf{g}}^{[i]}$ has a zero mean distribution and its covariance matrix can be computed as $Cov(\tilde{\mathbf{g}}^{[i]}) = \mathbb{E}(\tilde{\mathbf{g}}^{[i]H}\tilde{\mathbf{g}}^{[i]}) = \sigma_{BU}^2\mathbf{I} + \mathbf{\Omega}$, where $\mathbf{\Omega}$ is defined as

$$\mathbf{\Omega} = \mathbb{E}(\mathbf{G}_{BI}^H \operatorname{diag}(\tilde{\mathbf{g}}_{IU}^{[i]})^H \mathbf{f}^H \mathbf{f} \operatorname{diag}(\tilde{\mathbf{g}}_{IU}^{[i]})\mathbf{G}_{BI})$$

$$= \sum_{j=1}^{M} \begin{bmatrix} q_{j1}q_{j1}^*\mathbb{E}(h_j^{[i]}h_j^{[i]*}) & \cdots & q_{jN_T}q_{j1}^*\mathbb{E}(h_j^{[i]}h_j^{[i]*}) \\ \vdots & \ddots & \vdots \\ q_{j1}q_{jN_T}^*\mathbb{E}(h_j^{[i]}h_j^{[i]*}) & \cdots & q_{jN_T}q_{jN_T}^*\mathbb{E}(h_j^{[i]}h_j^{[i]*}) \end{bmatrix}$$

$$\approx \begin{bmatrix} \mathbb{E}(\mathbf{q}_1^H\mathbf{q}_1)\sigma_{IU}^2 & \cdots & \mathbb{E}(\mathbf{q}_{N_T}^H\mathbf{q}_1)\sigma_{IU}^2 \\ \vdots & \ddots & \vdots \\ \mathbb{E}(\mathbf{q}_1^H\mathbf{q}_{N_T})\sigma_{IU}^2 & \cdots & \mathbb{E}(\mathbf{q}_{N_T}^H\mathbf{q}_{N_T})\sigma_{IU}^2 \end{bmatrix}$$

$$= M\beta_{BI}\sigma_{IU}^2\mathbf{I} \qquad (31)$$

The notations $q_{ij}$, $h_j^{[i]}$ and $\mathbf{q}_j$ in (31) stand for the element in the $i$th row and the $j$th column of $\mathbf{G}_{BI}$, the $j$-th element of the vector $\tilde{\mathbf{g}}_{IU}^{[i]}$, and the $j$th column of $\mathbf{G}_{BI}$, respectively. Note that the approximation stands for large values of $M$.